\documentclass[12pt]{article}
\usepackage{amsfonts}
\usepackage[dvips]{graphicx}
\usepackage{latexsym,amsmath,amssymb,graphics,stmaryrd}
\usepackage{cite}
\usepackage{array}
\usepackage{subfigure}
\usepackage{multirow}
\usepackage{epsfig}
\usepackage[dvips]{graphicx}
\usepackage[dvips]{color}
\usepackage{pstricks}
\usepackage{pst-node}
\usepackage{pst-plot}
\usepackage{dsfont}
\usepackage{amsthm}
\usepackage{graphics}
\usepackage{subfigure}
\usepackage{fancyhdr}
\usepackage[dvips]{color}

\fancypagestyle{plain}{
\fancyhf{}
\newcommand{\fpage}{\iffloatpage{}{\thepage}}
\fancyfoot[C]{\fpage}

}

\newcommand{\be}{\begin{eqnarray}}
\newcommand{\ee}{\end{eqnarray}}

\newcommand{\la}{\lambda}

\newcommand{\NN}{\mathcal{N}}

\newcommand{\ms}{\!-\!}
\newcommand{\ps}{\!+\!}
\newcommand{\arcsinh}{{\rm arcsinh}}

\newlength{\neglength}
\newlength{\diameter}

  \newcommand{\nn}{\nonumber}
  
 \newcommand{\sfrac}[2]{\mbox{$\frac{#1}{#2}$}}

\DeclareMathOperator{\sgn}{sgn}

\numberwithin{equation}{section}
\newlength{\unit}
\psset{xunit=\unit,yunit=\unit,runit=\unit}
\newlength{\linew}
\psset{linewidth=\linew}
\begin{document}

\begin{flushright}\footnotesize
\texttt{UUITP-21/12} 
\vspace{0.8cm}\end{flushright}

\begin{center}
{\Large\textbf{\mathversion{bold} $N^3$-behavior from 5D Yang-Mills theory}}

\vspace{1.5cm}

\textrm{J.~K\"all\'en, J.~A.~Minahan, A.~Nedelin and M.~Zabzine}
\vspace{8mm}

\textit{ Department of Physics and Astronomy\\ Uppsala University\\ Box 520\\
SE-751 20 Uppsala, Sweden}\\

\vspace{3mm}


\par\vspace{1cm}

\textbf{Abstract} \vspace{5mm}

\begin{minipage}{14cm}

In this  note we derive $N^3$-behavior at  large 't Hooft coupling for the free energy of 
 5D maximally supersymmetric  Yang-Mills theory on $S^5$.  We also consider a $Z_k$ quiver of this model, as well as a model with $M$ hypermultiplets in the fundamental representation.
  We compare the results to the supergravity description and comment on their relation. 
\end{minipage}

\end{center}

\vspace{0.5cm}


\newpage
\setcounter{page}{2}
\renewcommand{\thefootnote}{\arabic{footnote}}
\setcounter{footnote}{0}
\setcounter{equation}{0}


\section{Introduction}
\label{s-intro}

Recently there has been renewed interest in 6-dimensional $(2,0)$ superconformal theories. These theories
 do not admit a standard Lagrangian description, making it  difficult to study them directly. Much of our information about these theories comes through the  AdS/CFT correspondence, where the $(2,0)$ theories are conjectured to be dual to $M$-theory (or supergravity) on an $AdS_7 \times S^4$  background.  In particular, the supergravity dual reveals a mysterious $N^3$ dependence for the free-energy for the $(2,0)$ theories \cite{Klebanov:1996un,Henningson:1998gx}.

 The $(2,0)$ theory lives on the boundary of $AdS_7$, which  in its Lorentzian version  can be chosen to be  $R\times S^5$.  However, the Euclidean counterpart to this boundary  can have  $R$ compactified to  $S^1$.
 For the dual theory,   compactifying one Euclidean direction to $S^1$  reduces the $(2,0)$ theory  to 5-dimensional maximally supersymmetric Yang-Mills (SYM) theory. Recently it 
     has been suggested that the maximal 5D SYM theory contains all degrees of freedom of the $(2,0)$ theory, where  the Kaluza-Klein states from the $S^1$ are mapped to 
the      instantons of the 5D theory \cite{Douglas:2010iu, Lambert:2010iw} (see also \cite{Bolognesi:2011nh}).  Since the $N^3$ behavior remains in the supergravity dual after compactification, one might  expect to find some indication of this $N^3$ behavior in  5D SYM.
   
In this note we    consider the recent calculations of the $\NN=1$ SYM  partition function on $S^5$.   We show in the case where there is one adjoint hypermultiplet, which in the large radius limit has an enhanced $\NN=2$   supersymmetry\footnote{We will refer to this as an $\NN=2$ model, even though it is not clear that the theory on $S^5$ actually preserves 16 supersymmetries.},  that the free energy scales as $N^3$,
   agreeing with the expectation from  supergravity. However, if we take 
   the suggested identification of  $g_{YM}^2$ with the radius of $S^1$ we find  a  small mismatch with the $N^3$ coefficient.   We also consider a $Z_k$ quiver of the $\NN=2$ model which also exhibits $N^3$ behavior.  In order for the $k$ dependence of the matrix model calculation to agree with the corresponding supergravity calculation, there should be an additional factor of $k$ in the identification of the $S^1$ radius and $g_{YM}^2$.  Finally, we consider the free-energy for  $\NN=1$ models with $M$ hypermultiplets in the fundamental representation.   In this case the free-energy scales as $N^2$ for $M\le 2N$ in the strong-coupling limit.  If $M>2N$ then the strong-coupling limit is destabilized. 
  
 In a related paper \cite{Kim:2012av}, it was shown that $N^3$ behavior can arise from a different formulation of SYM on $S^5$.  Here, localization reduces the partition function to one almost identical to a  Chern-Simons partition function, where it was previously demonstrated to have $N^3$ behavior in the strong-coupling limit \cite{Gopakumar:1998ki, Marino:2004eq, Marino:2005sj}.
 
The rest of this  note is organized as follows: in section \ref{s-5D-YM}  we briefly review  the structure of the partition function 
 for  5D SYM on $S^5$.  In section  \ref{s-matrix} we analyze the large $N$-behavior at large 
  't Hooft coupling of the corresponding matrix model for the $\NN=2$, its $Z_k$ quiver and  models with $M$ hypermultiplets in the fundamental representation. In section \ref{s-sugra} we review the supergravity analysis for 
   $AdS_7 \times S^4$. In section \ref{s-end} we compare the gauge theory result with the supergravity result
    and  comment on the numerical mismatch. 

\section{5D supersymmetric Yang-Mills theory on $S^5$}
\label{s-5D-YM}

In this section we briefly review  the status of 5D SYM theory on $S^5$. On $\mathbb{R}^5$ 
 the $\NN=1$  SYM theory is invariant under 8 supercharges while $\NN=2$ SYM  
  is maximally supersymmetric and is invariant under 16 supercharges.  The matter content of the $\NN=2$ theory contains an 
   $\NN=1$ vector multiplet plus an $\NN=1$ hypermultiplet in the adjoint representation. Recently in \cite{Hosomichi:2012ek}, $\NN=1$ supersymmetric Yang-Mills 
    with hypermultiplets has been constructed on $S^5$.  Since 5D SYM theory is not superconformal, there is no 
     canonical way to put it on $S^5$. However we can think of $\NN=1$ Yang-Mills theory with hypermultiplets as a deformation 
      of the flat theory controlled by the parameter $r$, where $r$ is the radius of $S^5$.  Once the limit $r\rightarrow \infty$ is taken,
       all formulae consistently collapse to the flat case. Thus,  $\NN=1$ SYM with a hypermultiplet in the adjoint representation on $S^5$ 
        produces a deformation of flat $\NN=2$ SYM, where 8 supercharges are explicitly preserved. 
    
  The partition function  is obtained using localization.    For the localization to work on $S^5$ one needs at least $\NN=1$ supersymmetry. Based on the earlier papers \cite{Kallen:2012cs} and
     \cite{Hosomichi:2012ek}, the localization for $\NN=1$ SYM was analyzed in  \cite{Kallen:2012va}.    There it was argued  that the full partition function for $\NN=1$ SYM theory with a hypermultiplet in representation $R$ has the following form
          \be
Z&=&\int\limits_{\rm Cartan} [d\phi]~e^{-  \frac{8\pi^3 r}{g_{YM}^2}  \text{Tr}(\phi^2)} {\rm det}_{\rm Ad}\left (  \sin ( i\pi  \phi) e^{ \frac{1}{2} f(i \phi )}  \right ) \nn \\
&&\times~  {\rm det}_{R} \left ( \left ( \cos ( i\pi \phi )\right )^{\frac{1}{4}} e^{-\frac{1}{4} f \left (\frac{1}{2} -  i\phi \right ) - \frac{1}{4} f \left (\frac{1}{2} +  i\phi \right )} \right )  + \mathcal{O} (e^{-\frac{16 \pi^3 r}{g_{YM}^2}})~,\label{vector1loop-intro}
\ee
 where $g_{YM}$ is the Yang-Mills coupling constant.  For the case of a hypermultiplet in the adjoint  representation the answer can be rewritten  in the following form
\begin{eqnarray}
\nonumber
Z=\int\limits_{Cartan}\left[d\phi\right]e^{-\frac{8\pi^3 r}{g_{YM}^2}\mathrm{Tr}(\phi^2)}\prod_{\beta}(\sin(\pi\langle\beta, i\phi\rangle) (\cos(\pi\langle\beta, i\phi\rangle))^{\frac{1}{4}}
\times\\
e^{\frac{1}{2}f(\langle\beta, i\phi\rangle)-
\frac{1}{4}f(\frac{1}{2}-\langle\beta, i\phi\rangle)-
\frac{1}{4}f(\frac{1}{2}+\langle\beta, i\phi\rangle)} +  {\mathcal{O}} (e^{-\frac{16 \pi^3 r}{g_{YM}^2}})~,\label{main-matrix}
\end{eqnarray}
where $\beta$ are the roots and $r$ is the radius of $S^5$. 
 Here the function $f(x)$  is given by the following expression 

\be
f(y)=\frac{i\pi y^3}{3}+y^2 \log\left(1-e^{-2\pi i y}\right)+\frac{i y}{\pi}\mathrm{Li}_{2}\left(e^{-2\pi i y}\right)
+\frac{1}{2\pi^2}\mathrm{Li}_{3}\left(e^{-2\pi i y}\right)-\frac{\zeta(3)}{2\pi^2}~.
\label{f:function}
\ee
 A very important property of (\ref{vector1loop-intro}) is that $\det_{R} (\cdots) = \sqrt {\det_{R} (\cdots)\det_{\bar{R}} (\cdots)}$ (see
  \cite{Kallen:2012va} for further explanation). 
 The matrix models in (\ref{vector1loop-intro}) and  (\ref{main-matrix})  correspond
  to the full perturbative partition functions (i.e.\ localization around the trivial connection).
 All corrections coming from instantons are contributing in with overall factors $\exp(-\frac{16 \pi^3 r}{g_{YM}^2})$, as was argued in \cite{Kallen:2012va}. 
   If we introduce the 't Hooft coupling constant 
$$\lambda=\frac{g_{YM}^2 N}{r}~,$$
 and consider the large $N$-limit of the partition function (\ref{main-matrix}) while keeping $\lambda$ fixed then 
  only the matrix integral (\ref{main-matrix}) contributes to the leading large $N$ behavior. The instanton contributions are exponentially 
   suppressed in the large $N$-limit with fixed  't Hooft coupling. 

  In \cite{Kim:2012av} the authors claim that one can construct an $\NN=2$ supersymmetric Yang-Mills theory on $S^5$ that  preserves 16 supercharges. 
   The model they consider belongs to a class of $\NN=1$ theories with hypermultiplets which admit a two parameter deformation controlled by the radius $r$ and a real parameter $\Delta$, very much in 
the spirit of the 3D story \cite{Jafferis:2010un}, although in the 5d case we believe that the reality properties of the one-loop determinants coming from the hypermultiplets need to be checked for generic values of $\Delta$.    Nevertheless,   with a single adjoint hypermultiplet, $\Delta=1$ corresponds to the $\NN=2$ model studied in  \cite{Kim:2012av}, while $\Delta=1/2$ corresponds to the model studied here.   As far as we can see, the correct deformation of $\NN=2$ supersymmetric 
     Yang-Mills theory associated to $S^5$ remains an open problem and requires further study. 
       Our main goal is to study the matrix models (\ref{vector1loop-intro}) and 
        (\ref{main-matrix}), but in the last section we will comment on the $N^3$ behavior for generic values of $\Delta$.

\section{$N^3$-behavior from the matrix model}
\label{s-matrix}

In this section we analyze the large $N$-behavior of the matrix models in (\ref{vector1loop-intro}) and (\ref{main-matrix}). 
 We explicitly find  $N^3$ scaling  for 
 the free-energy at the large 't Hooft coupling for the case of an adjoint hypermultiplet and $Z_k$ quiver theory. 
  We also consider the matrix model obtained from the minimal $\NN=1$ theory with $M$ hypermultiplets in the fundamental representation.  Here we show that the free-energy scales as $N^2$ for $M\le 2N$.  If $M>2N$ then the matrix model destabilizes in the strong-coupling limit.

We start with the matrix model  (\ref{main-matrix}) rewritten in terms of $\phi$ eigenvalues
\begin{eqnarray}
\nonumber
& &Z \sim  \int \prod_{i=1}^{N}d\phi_{i}\exp\left(-\frac{ 8\pi^3 r}{g_{YM}^2}\sum\limits_{i}\phi_{i}^2+
\sum\limits_{j\neq i}\sum\limits_{i}\left[\log\left[\sinh(\pi(\phi_i-\phi_j))\right]+
\frac{1}{4}\log\left[\cosh(\pi(\phi_i-\phi_j))\right]
\right.\right.
\\
& &\left.\left.
+ \frac{1}{2}f(i(\phi_i-\phi_j))-
\frac{1}{4}f\left(\frac{1}{2}+i(\phi_i-\phi_j)\right)-
\frac{1}{4}f\left(\frac{1}{2}-i(\phi_i-\phi_j)\right)\right]\right)~. 
\label{partition:function}
\end{eqnarray}
The derivative of the function $f(y)$ has the remarkably simple form,
\be
\frac{d f(y)}{dy}=\pi y^2 \cot(\pi y)\,.
\label{f:derivative}
\ee
Using this and some simple trigonometric identities we can derive the saddle point
equation for (\ref{partition:function}),
\begin{eqnarray}
\nonumber
\frac{16\pi^3 N}{\lambda}\phi_i&=& \pi \sum\limits_{j\neq i}\Big[\left(2- (\phi_i-\phi_j)^2\right)\coth(\pi(\phi_i-\phi_j))+
\\
&&\qquad\qquad\left(\frac{1}{4}+(\phi_i-\phi_j)^2\right)\tanh(\pi(\phi_i-\phi_j))\Big]\,,
\label{eom}
\end{eqnarray}
where we have introduced the 't Hooft coupling constant $\lambda=g_{YM}^2N/r$. In the
strong coupling  limit $\lambda\to\infty$  the eigenvalues are pushed apart and  the partition function (\ref{partition:function}) and equation of motion (\ref{eom}) can be approximated as
\be
Z\sim\int\prod_{i} d\phi_{i} e^{-\frac{8 \pi^3N}{\lambda}\sum\limits_{i}\phi_{i}^2+\frac{9\pi}{8}
\sum\limits_{j\neq i}\sum\limits_{i}|\phi_i-\phi_j|}
\label{partition:strong}
\ee
and
\be
\frac{16\pi^2 N}{\lambda}\phi_i=\frac{9}{4}\sum\limits_{j\neq i}\mathrm{sign}(\phi_i-\phi_j)\,,
\label{eom:strong}
\ee
respectively.
Assuming that the eigenvalues $\phi_i$ are ordered, we get the solution
\be
\phi_i=\frac{9\lambda}{ 64 \pi^2 N}(2i-N)~.
\label{solution:strong}
\ee
Taking the limit $N\to\infty$ and substituting the saddle point solution (\ref{solution:strong}) back into (\ref{partition:strong}), 
we find the free-energy,
\be
 F \equiv -\log Z\approx -\frac{27}{512 }\frac{g_{YM}^2 N^3}{\pi r}\,,\label{matrix-final}
\ee
 where we  used the approximations
\be
\sum\limits_{i=1}^{N}(2i-N)^2\approx \frac{1}{3}N^3\,,\qquad
\sum\limits_{j\neq i}\sum\limits_{i=1}^{N}|i-j|\approx \frac{1}{3}N^3\,.
\ee

A related theory to the $\NN=2$ model is a $Z_k$ quiver, where the  $SU(N)$ gauge group is broken to $SU(N/k)^k$ and with the hypermultiplets  in the bifundamental representations,
$(N/k,\overline{N/k},1,\dots 1)$, $(1,N/k,\overline{N/k},1,\dots )$, {\it etc.}.  The $N$ eigenvalues that appear in (\ref{eom}) can be split into $k$ groups of $N/k$, $\psi^{(r)}_i$, where $r=1,\dots,k$ and $i=1,\dots N/k$.  The equation of motion from the resulting matrix model 
  (\ref{vector1loop-intro}) is then
\begin{eqnarray}
\nonumber
\frac{16 \pi^3 N}{\lambda}\psi^{(r)}_i&=& \pi \Big[\sum\limits_{j\neq i}\left(2- (\psi^{(r)}_i-\psi^{(r)}_j)^2\right)\coth(\pi(\psi^{(r)}_i-\psi^{(r)}_j))
\\
&&\qquad+\sum_j\Big[\sfrac12\left(\sfrac{1}{4}+(\psi^{(r)}_i\ms\psi^{(r\ps1)}_j)^2\right)\tanh(\pi(\psi^{(r)}_i\ms\psi^{(r\ps1)}_j))\nn\\
&&\qquad\qquad+\sfrac12\left(\sfrac{1}{4}+(\psi^{(r)}_i\ms\psi^{(r\ms1)}_j)^2\right)\tanh(\pi(\psi^{(r)}_i\ms\psi^{(r\ms1)}_j))\Big]\Big]~.
\label{eom-quiver}
\end{eqnarray}
This has a solution where $\psi^{(r)}_i=\psi^{(s)}_i$, in which case (\ref{eom-quiver}) takes the same form as (\ref{eom}), except with $N$ replaced by $N/k$ in the summation limits.  Thus,  in the strong-coupling limit we have
\be
\phi_i=\frac{9g_{YM}^2 }{ 64 \pi^2 r}(2i-N/k)\,,
\label{solution:quiver}
\ee
with free-energy 
\be
 F \approx -k\frac{27}{ 512 }\frac{g_{YM}^2 N^3}{\pi k^3r}=-\frac{27}{ 512 }\frac{g_{YM}^2 N^3}{\pi k^2r}\,.\label{quiver-final}
 \ee

In these models, the $N^3$ behavior arises from the long-range linear repulsive potential between the eigenvalues. In fact, any matrix model with such a potential will give $N^3$ behavior since it will spread the eigenvalues over a range of order $N$.  
However,   a generic $\NN=1$ model will not have such a potential.  

For example, suppose we consider $M$ hypermultiplets in the fundamental and anti-fundamental representations.  In this case  the eigenvalue equation for matrix model (\ref{vector1loop-intro}) becomes
\begin{eqnarray}
\nonumber
\frac{16 \pi^3 N}{\lambda}\phi_i= \pi\Big( \sum\limits_{j\neq i}\left[\left(2- (\phi_i-\phi_j)^2\right)\coth(\pi(\phi_i-\phi_j))\right]
\\
+\frac{M}{2}\left(\frac{1}{4}+\phi_i^2\right)\tanh(\pi\phi_i)\Big)\,.
\label{eom2}
\end{eqnarray}
Since $N^3$ behavior requires well separated eigenvalues, let us assume that they are, in which case we can approximate (\ref{eom2}) as 
\begin{eqnarray}
\nonumber
\frac{16 \pi^3 N}{\lambda}\phi_i= \pi\Big( \sum\limits_{j\neq i}\left[\left(2- (\phi_i-\phi_j)^2\right)\sgn(\phi_i-\phi_j)\right]
\\
+\frac{M}{2}\left(\frac{1}{4}+\phi_i^2\right)\sgn(\phi_i)\Big)\,.
\label{eom3}
\end{eqnarray}
Taking the limit $N\to\infty$ and defining $x=i/N\ms1/2$, we can rewrite (\ref{eom3}) as
\be\label{eom4}
\frac{16 \pi^3 N}{\lambda}\phi(x)&=&N\pi\Big [4x\ms2\Big(x\phi^2(x)\ms2\phi(x)\left(\Phi(x)\ms\Phi(\sfrac12)\right)+\Phi_2(x)\Big)\nn\\
&&\qquad\qquad+ \frac{M}{2N}\left(\frac14+\phi^2(x)\right)\sgn(x)\Big]~,
\ee
where $\Phi(x)=\int_0^x \phi\, dx$ and $\Phi_2(x)=\int_0^x \phi^2 dx$.  In (\ref{eom4}) we have assumed that $\phi(x)$ is a monotonic increasing odd function, although not necessarily continuous.  

If we now take an $x$ derivative on both sides of (\ref{eom4}), we end up with the equation
\be\label{eom5}
\frac{ 16 \pi^3 }{\lambda}\phi'(x)&=&4\pi \Big [1-(x- b)\phi(x)\phi'(x)+\phi'(x)\left(\Phi(x)\ms\Phi(\sfrac12)\right)\Big ]~,\quad x>0\nn\\
\frac{16\pi^3 }{\lambda}\phi'(x)&=&4\pi \Big [1-(x+ b)\phi(x)\phi'(x)+\phi'(x)\left(\Phi(x)\ms\Phi(\sfrac12)\right ) \Big ]~,\quad x<0\nn\\
\ee
where $b=M/(4N)$.  Dividing by $\phi'(x)$ and taking one more $x$ derivative, we arrive at
\be
0&=&-\frac{\phi''}{(\phi')^2}-(x-b)\phi'\,,\qquad x>0\nn\\
0&=&-\frac{\phi''}{(\phi')^2}-(x+b)\phi'\,,\qquad x<0
\ee
which has the solution 
\be\label{phisol}
\phi(x)&=&\arcsinh\left(\frac{x-b}{c}\right)+C\,,\qquad 0<x\le\sfrac{1}{2}\nn\\
\phi(x)&=&\arcsinh\left(\frac{x+b}{c}\right)-C\,,\qquad -\sfrac{1}{2}\le x<0\,,
\ee
where $c$ and $C$ are  yet to be determined constants.  Substituting this solution back into (\ref{eom5}) leads to the relation
\be\label{sol1}
\frac{8 \pi^2}{\la}=\sqrt{4c^2+(1-2b)^2}-(1-2b)\left(\arcsinh\left(\frac{1-2b}{2c}\right)+C\right)\,.
\ee
If we substitute (\ref{phisol}) into (\ref{eom4}), then using (\ref{sol1}) we find the solutions for  $C$
\be\label{sol2}
C=\arcsinh\frac{b}{c}\pm\sqrt{\frac{c^2}{b^2}+\frac{5}{4}}-\sqrt{\frac{c^2}{b^2}+1}\,,
\ee
where only the solution with the $+$ sign is consistent with the monotonicity of $\phi(x)$.

If $b<1/2$ then there exists a positive real value of $c$ that satisfies (\ref{sol1}) and (\ref{sol2}), even if $\la\to\infty$.  (\ref{sol2})  shows that $\phi(x)$ is positive (negative) for positive (negative) $x$ and has a finite jump at $x=0$.  Since $c$ is nonzero, (\ref{phisol}) and (\ref{sol2}) show that the eigenvalues are distributed over a finite range and thus the approximation in (\ref{eom3}) is not valid.  Nonetheless, it is still true that  the eigenvalues are only over a finite extent, hence the  free-energy can only scale as $N^2$ since all $\phi_i$ and $\phi_i-\phi_j$ are finite in the large $N$ limit.

If $b=1/2$, which corresponds to $M=2N$, then (\ref{sol1}) gives $c=4 \pi^2/{\la}$.  In the strong coupling limit, $c\to 0$ and we can approximate $\phi(x)$ as
\be\label{phisol2}
\phi(x)&\approx&\log\frac{1}{1-2x}+\frac{\sqrt{5}-2}{2}\qquad\qquad 0<x< \sfrac12,\nn\\
\phi(x)&\approx&-\log\frac{1}{1+2x}-\frac{\sqrt{5}-2}{2}\qquad -\sfrac12< x<0
\ee
away from the boundary points $x=\pm\sfrac{1}{2}$ and
\be
\phi(\pm\sfrac{1}{2})\approx \pm\left(\log\frac{1}{c}+\frac{\sqrt{5}-2}{2}\right)
\ee
at these points. Hence, the eigenvalues spread out over an infinite distance as $c\to0$.  However, a finite fraction are within a finite region, for example, half the eigenvalues lie between $\pm (\log 2+\sfrac{\sqrt{5}-2}{2})$.  Thus, the approximation in (\ref{eom3}) is not completely valid.  Using it anyway, one can easily check that the free-energy that gives the equation of motion in (\ref{eom2}) scales as $N^2$ with the eigenvalue distribution in  (\ref{phisol2}).

Finally, if $b>1/2$ then (\ref{sol1}) has no real solution for $c$ in the strong coupling limit, suggesting that the eigenvalue distribution destabilizes.

 \section{Comparison with supergravity on $AdS_7\times S^4$}
\label{s-sugra}

We now compare our results in the previous section to the supergravity result on $AdS_7\times S^4$ where the $AdS_7$ boundary is $S^1\times S^5$.   The radius of $AdS_7$ is $\ell$, while that of the $S^4$ is $\ell/2$, where $\ell=2\ell_{pl}(\pi N)^{1/3}$.  The $AdS_7$ metric can then be written in the form
\be
ds^2=\ell^2(\cosh^2\rho\, d\tau^2+d\rho^2+\sinh^2\rho\, d\Omega_5^2)~,
\ee
where $d\Omega_5^2$ is the round metric for the unit 5-sphere and $\tau\equiv \tau+2\pi R_6/r$.  $R_6$ and $r$ are the radii of $S^1$ and $S^5$ on the boundary.

According to the AdS/CFT correspondence, the supergravity classical action equals the free-energy of the boundary field theory.  The action itself is divergent so it needs to be regulated by adding counterterms \cite{Balasubramanian:1999re,Emparan:1999pm,deHaro:2000xn,Awad:2000aj}.  The full action then has the form
\be
I_{AdS}=I_{\rm{bulk}}+I_{\rm{surface}}+I_{\rm{ct}}\,,
\ee
where
\be
I_{\rm{bulk}}=-\frac{1}{16\pi G_N}\mbox{Vol}(S^4)\int d^7x \sqrt{g}\left(R-2\Lambda\right)\
\ee
is the action in the bulk, $I_{\rm{surf}}$ is the surface contribution and $I_{\rm{ct}}$ contains counterterms written only in terms of the boundary metric and which cancel off divergences in $I_{\rm{bulk}}$.  We use the convention in \cite{Maldacena:1997re} for $G_N$, $G_N={16\pi^7\ell_{pl}^9}$.  Using that
\be
R-2\Lambda=-\frac{12}{\ell^2}\,,
\ee
we have
 \be\label{bulk}
I_{\rm{bulk}}=-\frac{1}{256\pi^8 \ell_{pl}^9}\left(\frac{\pi^2\ell^4}{6}\right)\frac{2\pi R_6}{r}\pi^3 (-12\ell^5)\int_0^{\rho_0}\cosh\rho\sinh^5\rho\, d\rho=\frac{4\pi R_6}{3\,r} N^3\sinh^6\rho_0\,.\nn\\
\ee
In the limit that $\rho_0\to\infty$ the integral is divergent and corresponds to a UV divergence for the boundary theory.  In terms of an $\epsilon$ expansion of the boundary theory, we make the identification $\epsilon=e^{-\rho_0}$, which then gives
\be
\sinh^6\rho_0=\frac{1}{64}\epsilon^{-6}-\frac{3}{32}\epsilon^{-4}+\frac{15}{64}\epsilon^{-2}-\frac{5}{16}+{\rm O}(\epsilon^2)\,.
\ee
The surface term contributes to the divergent pieces, but not the finite part of (\ref{bulk}), while
the effect of the counterterm is to cancel off the divergent pieces.  Hence, we find \cite{Emparan:1999pm}
\be\label{IAdS}
I_{AdS}=-\frac{5\pi R_6}{12\,r}N^3~.\label{sugra-final}
\ee

The supergravity dual of a $(2,0)$ $Z_k$ quiver theory is expected to be $AdS_7\times S^4/Z_k$, where $S^4/Z_k$ is a $Z_k$ orbifold of $S^4$ \cite{Ferrara:1998vf}.  The only change in the preceding calculation is to replace $\mbox{Vol}(S^4)$ with the $\mbox{Vol}(S^4/Z_k)=\mbox{Vol}(S^4)/k$.  Hence, the regularized action is
\be
I_{AdS}=-\frac{5\pi R_6}{12\,k\,r}N^3~.\label{sugra-quiver}
\ee

\section{Discussion}\label{s-end}

We can now compare the gauge theory result in (\ref{matrix-final}) with the supergravity result in (\ref{sugra-final}).  
 The good news is that they both have  $N^3$ behavior. To compare the numerical factors 
 we need a relation between 
 $g_{YM}^2$ and $R_6$. As suggested in \cite{Douglas:2010iu, Lambert:2010iw} we can identify  the KK states on $S^1 \times \mathbb{R}^5$ 
   with the instanton particles on $\mathbb{R}^5$ and arrive at the following identification\footnote{Notice that we work with the following normalization of the Yang-Mills action $\frac{1}{2 g_{YM}^2} \int d^5x \sqrt{g}\, \mbox{Tr}F_{mn} F^{mn}$.} 
 \be
 R_6=\frac{g_{YM}^2}{8\pi^2}~.\label{relation}
 \ee
  Using this relation the supergravity result becomes
\be
I_{AdS}=-\frac{5 \,g_{YM}^2}{96\pi\,r}N^3\,,
\ee
which is off by a factor of ${81}/{80}$ from the gauge theory calculation (\ref{matrix-final}). 
For the quiver theory, if we use (\ref{relation}) then the power of $k$ in (\ref{quiver-final}) does not match with (\ref{sugra-quiver}).  This suggests that the identification between the $S^1$ radius and $g_{YM}^2$ should be
 \be
 R_6=\frac{g_{YM}^2}{8\pi^2k}~.\label{relation-quiver}
 \ee

If we take the matrix model suggested in 
   \cite{Kim:2012av} (which  only has the sine factors in the determinant), then as pointed out in   \cite{Kim:2012av} one can evaluate the integral directly  \cite{Marino:2002fk, Tierz:2002jj}, where one finds a factor of $N(N^2-1)$ in the free-energy.  Alternatively, one can use  the analysis from section \ref{s-matrix} to find the leading $N^3$ factor.  The resulting free-energy is given by (\ref{matrix-final}) multiplied by a factor of $64/81$. This  still has the $N^3$-behavior,
   but the numerical mismatch with  (\ref{IAdS}) remains.  If we consider the more general models in \cite{Kim:2012av} parameterized by $\Delta$, then the analysis in section  \ref{s-matrix}  gives
\be\label{genDelta}
 F \equiv -\log Z\approx -\frac{(2-\Delta)^2(1+\Delta)^2}{96 }\frac{g_{YM}^2 N^3}{\pi r}\,,\qquad -1<\Delta<2\,,
\ee
which is minimized for $\Delta=1/2$.  If $\Delta$ is outside the bounds in (\ref{genDelta}) then there will be long-range attraction between the eigenvalues which cannot lead to $N^3$ behavior \cite{Marino:2012az}.

A possible explanation for the numerical mismatch is that    we are looking at the wrong 5D SYM theory on $S^5$. Since the theory is not superconformal, there is no 
 canonical way to put it on the sphere. Moreover, we can add to the 5D Yang-Mills action  a supersymmetric Chern-Simons term, thus modifying the numerics of the matrix model.  
  Another possibility is that the relation between $R_6$ and $g_{YM}$ on $S^1 \times S^5$ differs from the one suggested
 in (\ref{s-matrix}) by  \cite{Douglas:2010iu, Lambert:2010iw}.  {{We think that the relation between $(2,0)$ 6D theory
  and supersymmetric 5D Yang-Mills theory should be understood better. The results presented in this work can be used
   to actually check the different conjectures}}
 
 Finally,  the finite part of (\ref{bulk}) is actually scheme dependent, as one can add a local counterterm to the boundary which is proportional to the conformal anomaly \cite{deHaro:2000xn}\footnote{We thank Kostas Skenderis for pointing this out to us.}.  Choosing a different scheme could then change (\ref{IAdS}).  In fact, since we are really considering a 5-dimensional theory, it may be more appropriate to consider supergravity backgrounds sourced by D4 branes \cite{Kanitscheider:2008kd,Kanitscheider:2009as}.  In this case,  one could also allow local counterterms that are covariant in five dimensions but not in six  \cite{KS-private}.

\subsection*{Acknowledgments}
We thank Seok Kim, Silviu Pufu,  Kostas Skenderis, Jian Qiu and Konstantin Zarembo for correspondence and discussions.  This  research is supported in part by
Vetenskapsr\aa det under grants \#2009-4092 and \#2011-5079.  J.A.M thanks the
CTP at MIT   for kind
hospitality  during the course of this work.

\footnotesize

\begin{thebibliography}{10}

\bibitem{Klebanov:1996un}
I.~R. Klebanov and A.~A. Tseytlin, {\it {Entropy of Near Extremal Black
  P-Branes}},  {\em Nucl.Phys.} {\bf B475} (1996) 164--178,
  [\href{http://xxx.lanl.gov/abs/hep-th/9604089}{{\tt hep-th/9604089}}].

\bibitem{Henningson:1998gx}
M.~Henningson and K.~Skenderis, {\it {The Holographic Weyl Anomaly}},  {\em
  JHEP} {\bf 9807} (1998) 023,
  [\href{http://xxx.lanl.gov/abs/hep-th/9806087}{{\tt hep-th/9806087}}].

\bibitem{Douglas:2010iu}
M.~R. Douglas, {\it {On D=5 super Yang-Mills theory and (2,0) theory}},  {\em
  JHEP} {\bf 1102} (2011) 011, [\href{http://xxx.lanl.gov/abs/1012.2880}{{\tt
  arXiv:1012.2880}}].

\bibitem{Lambert:2010iw}
N.~Lambert, C.~Papageorgakis, and M.~Schmidt-Sommerfeld, {\it {M5-Branes,
  D4-Branes and Quantum 5D Super-Yang-Mills}},  {\em JHEP} {\bf 1101} (2011)
  083, [\href{http://xxx.lanl.gov/abs/1012.2882}{{\tt arXiv:1012.2882}}].

\bibitem{Bolognesi:2011nh}
S.~Bolognesi and K.~Lee, {\it {Instanton Partons in 5-dim SU(N) Gauge Theory}},
   {\em Phys.Rev.} {\bf D84} (2011) 106001,
  [\href{http://xxx.lanl.gov/abs/1106.3664}{{\tt arXiv:1106.3664}}].

\bibitem{Kim:2012av}
H.-C. Kim and S.~Kim, {\it {M5-branes from gauge theories on the 5-sphere}},
  \href{http://xxx.lanl.gov/abs/1206.6339}{{\tt arXiv:1206.6339}}.

\bibitem{Gopakumar:1998ki}
R.~Gopakumar and C.~Vafa, {\it {On the Gauge Theory / Geometry
  Correspondence}},  {\em Adv.Theor.Math.Phys.} {\bf 3} (1999) 1415--1443,
  [\href{http://xxx.lanl.gov/abs/hep-th/9811131}{{\tt hep-th/9811131}}].

\bibitem{Marino:2004eq}
M.~Mari\~no, {\it {Les Houches Lectures on Matrix Models and Topological
  Strings}},  \href{http://xxx.lanl.gov/abs/hep-th/0410165}{{\tt
  hep-th/0410165}}.

\bibitem{Marino:2005sj}
M.~Mari\~no, {\it {Chern-Simons Theory, Matrix Models, and Topological
  Strings}}, International Series of Monographs on Physics, {\bf 131}.
  {\it  The Clarendon Press,  Oxford University Press, Oxford}, 2005.

\bibitem{Hosomichi:2012ek}
K.~Hosomichi, R.-K. Seong, and S.~Terashima, {\it {Supersymmetric Gauge
  Theories on the Five-Sphere}},  Nucl.\ Phys.\ B {\bf 865} (2012) 376
  [\href{http://xxx.lanl.gov/abs/1203.0371}{{\tt
  arXiv:1203.0371}}].

\bibitem{Kallen:2012cs}
J.~K\"all\'en and M.~Zabzine, {\it {Twisted supersymmetric 5D Yang-Mills theory and
  contact geometry}},  {\em JHEP} {\bf 1205} (2012) 125,
  [\href{http://xxx.lanl.gov/abs/1202.1956}{{\tt arXiv:1202.1956}}].

\bibitem{Kallen:2012va}
J.~K\"all\'en, J.~Qiu, and M.~Zabzine, {\it {The perturbative partition function of
  supersymmetric 5D Yang-Mills theory with matter on the five-sphere}},
    {\em JHEP} {\bf 1208} (2012) 157,
  [\href{http://xxx.lanl.gov/abs/1206.6008}{{\tt arXiv:1206.6008}}].

\bibitem{Jafferis:2010un}
D.~L. Jafferis, {\it {The Exact Superconformal R-Symmetry Extremizes Z}},  {\em
  JHEP} {\bf 1205} (2012) 159, [\href{http://xxx.lanl.gov/abs/1012.3210}{{\tt
  arXiv:1012.3210}}].

\bibitem{Balasubramanian:1999re}
V.~Balasubramanian and P.~Kraus, {\it {A Stress Tensor for Anti-de~Sitter
  Gravity}},  {\em Commun.Math.Phys.} {\bf 208} (1999) 413--428,
  [\href{http://xxx.lanl.gov/abs/hep-th/9902121}{{\tt hep-th/9902121}}].

\bibitem{Emparan:1999pm}
R.~Emparan, C.~V. Johnson, and R.~C. Myers, {\it {Surface Terms as Counterterms
  in the AdS / CFT Correspondence}},  {\em Phys.Rev.} {\bf D60} (1999) 104001,
  [\href{http://xxx.lanl.gov/abs/hep-th/9903238}{{\tt hep-th/9903238}}].

\bibitem{deHaro:2000xn}
S.~de~Haro, S.~N. Solodukhin, and K.~Skenderis, {\it {Holographic
  Reconstruction of Space-Time and Renormalization in the AdS / CFT
  Correspondence}},  {\em Commun.Math.Phys.} {\bf 217} (2001) 595--622,
  [\href{http://xxx.lanl.gov/abs/hep-th/0002230}{{\tt hep-th/0002230}}].

\bibitem{Awad:2000aj}
A.~M. Awad and C.~V. Johnson, {\it {Higher Dimensional Kerr - AdS Black Holes
  and the AdS / CFT Correspondence}},  {\em Phys.Rev.} {\bf D63} (2001) 124023,
  [\href{http://xxx.lanl.gov/abs/hep-th/0008211}{{\tt hep-th/0008211}}].

\bibitem{Maldacena:1997re}
J.~M. Maldacena, {\it {The Large $N$ Limit of Superconformal Field Theories and
  Supergravity}},  {\em Adv. Theor. Math. Phys.} {\bf 2} (1998) 231--252,
  [\href{http://xxx.lanl.gov/abs/hep-th/9711200}{{\tt hep-th/9711200}}].

\bibitem{Ferrara:1998vf}
S.~Ferrara, A.~Kehagias, H.~Partouche, and A.~Zaffaroni, {\it {Membranes and
  Five-Branes with Lower Supersymmetry and Their AdS Supergravity Duals}},
  {\em Phys.Lett.} {\bf B431} (1998) 42--48,
  [\href{http://xxx.lanl.gov/abs/hep-th/9803109}{{\tt hep-th/9803109}}].

\bibitem{Marino:2002fk}
M.~Mari\~no, {\it {Chern-Simons Theory, Matrix Integrals, and Perturbative
  Three Manifold Invariants}},  {\em Commun.Math.Phys.} {\bf 253} (2004)
  25--49, [\href{http://xxx.lanl.gov/abs/hep-th/0207096}{{\tt
  hep-th/0207096}}].
  
  \bibitem{Tierz:2002jj} 
  M.~Tierz,
  {\it {Soft matrix models and Chern-Simons partition functions}},
  Mod.\ Phys.\ Lett.\ A {\bf 19}, 1365 (2004),
[\href{http://xxx.lanl.gov/abs/hep-th/0212128}{{\tt
  hep-th/0212128}}].  
  

\bibitem{Marino:2012az}
M.~Mari\~no and P.~Putrov, {\it {Interacting Fermions and ${\mathcal{N}}\!=2$
  Chern-Simons-Matter Theories}},
  \href{http://xxx.lanl.gov/abs/1206.6346}{{\tt arXiv:1206.6346}}.

\bibitem{Kanitscheider:2008kd}
I.~Kanitscheider, K.~Skenderis, and M.~Taylor, {\it {Precision Holography for
  Non-Conformal Branes}},  {\em JHEP} {\bf 0809} (2008) 094,
  [\href{http://xxx.lanl.gov/abs/0807.3324}{{\tt arXiv:0807.3324}}].

\bibitem{Kanitscheider:2009as}
I.~Kanitscheider and K.~Skenderis, {\it {Universal Hydrodynamics of
  Non-Conformal Branes}},  {\em JHEP} {\bf 0904} (2009) 062,
  [\href{http://xxx.lanl.gov/abs/0901.1487}{{\tt arXiv:0901.1487}}].

\bibitem{KS-private}
K.~Skenderis. {\it Private communication}.

\end{thebibliography}

\providecommand{\href}[2]{#2}\begingroup\raggedright\endgroup

\end{document}